\documentclass{article}

\usepackage{amssymb}
\usepackage{epsfig}
\usepackage{color}
\usepackage{amsfonts}
\usepackage{amsmath}

\newcommand{\spacer}{\rule[0cm]{0cm}{0cm}}

\newcounter{statementnumber}
\renewcommand{\thestatementnumber}{\arabic{section}.\arabic{statementnumber}}

\newenvironment{thm}[2]{\refstepcounter{statementnumber} \label{#2}
\par \noindent {\bf #1~\thestatementnumber.}  
\begin{em}}{\end{em} \par}

\newenvironment{rem}[2]{\refstepcounter{statementnumber} \label{#2}
\par \noindent {\bf #1~\thestatementnumber.}
}{\par}

\newenvironment{proof}{\par\noindent\textsc{Proof.}}{\nopagebreak\spacer\hfill $\square$}

\newcounter{figurecount}

\def\mb{$\begin{displaystyle}}
\def\me{\end{displaystyle}$\ }

\def\rank{\mbox{\rm rank}\;}
\newcommand{\ket}[1]{\left| #1 \right\rangle}
\newcommand{\bra}[1]{\left\langle #1 \right|}
\newcommand{\inprod}[2]{\left\langle #1 | #2 \right\rangle}

\newcommand{\twotwo}[4]{\left[ \begin{array}{cc} #1 & #2 \\
                        #3 & #4 \end{array} \right]}

\newcommand{\real}{\mbox{\rm Re}}
\newcommand{\imag}{\mbox{\rm Im}}

\newcommand{\R}{{\mathbb R}}

\newcommand{\Proj}{{\mathbb P}}
\newcommand{\C}{{\mathbb C}}

\newcommand{\Z}{{\mathbb Z}}

\newcommand{\SU}{{\rm SU}}
\newcommand{\su}{{\rm su}}
\newcommand{\SO}{{\rm SO}}

\begin{document}

\thispagestyle{empty}

\begin{center}
{\Large \bf Minimum orbit dimension\\ \vspace*{-.5ex}for local unitary action\\ \vspace*{.2ex}on
$n$-qubit pure states}\\

\medskip

David W. Lyons\\
lyons@lvc.edu\\
Mathematical Sciences\\
Lebanon Valley College\\
\medskip
Scott N. Walck\\
walck@lvc.edu\\
Department of Physics\\
Lebanon Valley College\\
\medskip
revised: 9 September 2005\\
\end{center}


\thispagestyle{empty}




\begin{center}
\begin{minipage}{4in}
{\bf Abstract}. The group of local unitary transformations partitions
the space of $n$-qubit quantum states into orbits, each of which is a
differentiable manifold of some dimension.  We prove that all orbits of
the $n$-qubit quantum state space have dimension greater than or equal
to $3 n / 2$ for $n$ even and greater than or equal to $(3 n + 1)/2$ for
$n$ odd.  This lower bound on orbit dimension is sharp, since $n$-qubit
states composed of products of singlets achieve these lowest orbit
dimensions.
\end{minipage}
\end{center}

\setcounter{statementnumber}{0}
\section{Introduction}

Quantum entanglement theory can be regarded as the branch of
nonrelativistic quantum mechanics that seeks to understand the states
and dynamics of composite quantum systems with a fixed number of
subsystems.  Composite quantum systems can exhibit correlations among
subsystems in ways that classically describable composite systems
cannot.  A (pure) state of a composite quantum system is called
entangled if it cannot be described by specifying (pure) states for each
of the subsystems.

Quantum entanglement plays a particularly important role in quantum
information, where the subsystems are quantum bits or qubits (a
spin-$1/2$ particle is a physical realization of a qubit).  An $n$-qubit
system is the quantum analog of an $n$-bit computer or communications
channel.  Because quantum computing algorithms and quantum
communications protocols utilize entanglement as an essential resource,
potential applications of quantum information theory provide motivation
for a more complete description of entanglement
(see~\cite{nielsenchuang,gudder03} for surveys of a broad range of
topics in this area).

A fundamental problem in the theory of quantum entanglement is to
describe the types of entanglement that are achievable for a composite
quantum system.  We regard two states of a composite quantum system as
having the same type of entanglement if unitary operations on the
subsystems, called local unitary or LU transformations, can transform
one quantum state into the other.  Local unitary transformations form a
Lie group which acts on the manifold of quantum states, partitioning it
into orbits.  Each orbit is a differentiable manifold that represents a
type of quantum entanglement.  The \emph{orbit space}---the set of
orbits made into a topological space by the quotient topology---is the
collection of entanglement types.

A theory of quantum entanglement based on local unitary transformations
seeks to describe the orbit spaces and the orbits themselves for
composite quantum systems.  Much of the progress toward understanding
the orbit spaces of quantum systems comes from invariant theory---the
study of functions which are constant along orbits~\cite{rains00,
grassl98, sudbery01,makhlin02,meyer02,brylinski02a,brylinski02,
meyer01p,albeverio03,leifer04}.  One hopes to use these invariants,
which are usually polynomial functions of state vector coefficients, to
distinguish and classify orbits.  Rains~\cite{rains00} and Grassl et
al.~\cite{grassl98} laid the groundwork for a systematic
approach using this philosophy.  The success in choosing particular,
finite sets of invariants to label points in the orbit space has so far
been limited to small numbers of qubits.  Makhlin~\cite{makhlin02} gave
a set of 18 polynomial invariants that separate orbits for two-qubit
mixed states.  Sudbery~\cite{sudbery01} gave a set of six polynomial
invariants that separate orbits for 3-qubit pure states.  Ac\'in et
al.~\cite{acin00,acin01} gave a convenient set of non-polynomial
invariants and a classification of 3-qubit pure states based on it.

In this paper we pursue a strategy inspired by Linden and
Popescu~\cite{linden98,linden99}, who approached entanglement properties
of quantum states working on the Lie algebra level to study the orbits
themselves.  We develop a general technique for calculating the orbit
dimension of a state and use this to prove a lower bound on orbit
dimension.  We have also used our methods to provide a
proof~\cite{lyonswalckmaxdim} of the authors' claim
in~\cite{linden98,linden99} that almost all states have orbit dimension
$3n$ (we take the manifold of pure $n$-qubit states to be the projective
space $\Proj\left((\C^2)^{\otimes n}\right)$ and the group of local
unitary transformations to be $G=\SU(2)^n$).

Most of the progress in understanding orbits and orbit dimensions has
been for systems of only for two or three qubits.  Carteret and
Sudbery~\cite{carteret00a} described the non-generic orbits (including
orbit dimensions) for pure 3-qubit states.  \.Zyczkowski et
al.~\cite{kus01,sinolecka02} analyze orbits for bipartite states.  Few
general results are known about those orbits which are the most
interesting from the quantum information point of view, namely the
non-generic or exceptional orbits of $n$-qubit states (basic examples
are the singlet state of two qubits and the GHZ state of three qubits).
The main result in the present paper is at least a small step towards
the larger goal of orbit classification for general $n$.

\subsection*{Physical Significance of the Result}

In this paper, we identify the minimum orbit dimension of $n$-qubit
quantum states.  States that have the minimum orbit dimension are, in
some sense, the ``rarest'' quantum states.  Until now, it has been known
that singlet states have minimum orbit dimension for two qubits, and one
could conjecture that some $n$-qubit generalization of the singlet state
would have minimum orbit dimension for $n$ qubits, but it was not clear
how the singlet should be generalized to maintain the minimum orbit
dimension as $n$ increases.  For example, one generalization of the
singlet is the so-called $n$-cat state, $(1/\sqrt{2}) \ket{00\cdots00} +
(1/\sqrt{2}) \ket{11\cdots11}$, of which the GHZ state is an example for
three qubits.  But the $n$-cat generalization of the singlet does not
maintain the minimum orbit dimension for higher qubit numbers.  As we
show in this paper, it is the product of singlet states (for even qubit
numbers) or the product of singlets and one unentangled qubit (for odd
qubit numbers) that is the generalization of singlets that achieves
minimum orbit dimension.  This suggests a special role for the 2-qubit
singlet state in the theory of $n$-qubit quantum entanglement.

\subsection*{Proof Strategy and Outline}
To establish the minimum orbit dimension, we show that the orbit
dimension of a given state is (one less than) the rank of a real matrix $M$
associated to that state.  The matrix $M$ arises
naturally via consideration of the action of the local unitary group on
an infinitesimal level, that is, the action of the Lie algebra of the
local unitary group.  The column vectors of $M$ can be identified
with complex vectors. We then establish lower
bounds on the rank of $M$ by showing that a sufficient number of real dot
products of columns of $M$ can be arranged, possibly after local unitary
operations, to vanish.  Instead of working directly with real dot
products, it is convenient to calculate complex inner products; the
vanishing of the real part of a complex inner product guarantees that
the real dot product is zero (see~(\ref{complexvsrealinnprod}) below).

In~\S\ref{liealgactsect} we introduce the matrix $M$.  To establish the
necessary cancellations among terms of complex inner products of columns
of $M$ requires careful bookkeeping and a technical lemma; we present
this machinery in~\S\ref{techlemmas}.  Next we establish orthogonality
among columns of $M$ in~\S\ref{orthressect} and~\S\ref{orthgensect}.  We
then use these results to prove minimum orbit dimension
in~\S\ref{minorbthm}.

\setcounter{statementnumber}{0}
\section{Conventions and notation}

\subsection*{Hilbert space, state space and the local unitary group}

Let $H= (\C^2)^{\otimes n}$ denote the Hilbert space of pure states of a
system of $n$ qubits and let $\Proj(H)$ denote the projectivization of
$H$ which is the state space of the system.  We take the local unitary
group to be $G=\SU(2)^n$.  These definitions constitute a minor
departure, made for the sake of clarity and compactness of exposition,
from the widespread practice of taking state space to be the set of
normalized state vectors and resolving phase ambiguity by including an
extra ${\rm U}(1)$ factor in the local unitary group.

\subsection*{Multi-index notation for Hilbert space basis vectors}

Let $|0\rangle$, $|1\rangle$ denote
the standard basis for $\C^2$ and write $|i_{1}i_{2}\ldots i_n\rangle$ for 
$|i_{1}\rangle\otimes|i_{2}\rangle\otimes\cdots\otimes|i_n\rangle$ in
$(\C^2)^{\otimes n}$.  For a multi-index $I=(i_{1}i_{2}\ldots i_n)$ with $i_k
= 0,1$ for $1\leq k\leq n$, we will write $|I\rangle$ to denote
$|i_{1}i_{2}\ldots i_n\rangle$.  Let $i_k^c$ denote the bit complement
$$i_k^c = \left\{\begin{array}{ll}
0 & \mbox{if $i_k = 1$}\\
1 & \mbox{if $i_k = 0$}
\end{array}\right.$$
and let $I_k$ denote the multi-index 
$$I_k:=(i_{1}i_{2}\ldots i_{k-1} i_k^c i_{k+1} \ldots i_n)$$ obtained from
$I$ by taking the complement of the $k$th bit for $1\leq k\leq n$.
Similarly, let $I_{kl}$ denote the multi-index 
$$I_{kl}:=(i_{1}i_{2}\ldots i_{k-1} i_k^c i_{k+1} \ldots i_{l-1} i_l^c
i_{l+1} \ldots i_n)$$ obtained from $I$ by taking the complement of the $k$th
and $l$th bits for $1\leq k<l\leq n$.

\subsection*{Standard identification of $\C^N$ with $\R^{2N}$}

We identify the complex vectors in $\C^N$ with real vectors in $\R^{2N}$ via
\begin{eqnarray}
\C^N &\leftrightarrow &\R^{2N} \nonumber \\
(z_1,z_2,\ldots,z_N) &\leftrightarrow &(a_1,b_1,a_2,b_2,\ldots,a_N,b_N) \label{cnidentr2n}
\end{eqnarray}
where $z_j = a_j + ib_j$ for $1\leq j\leq N$.

\setcounter{statementnumber}{0}
\section{Lie algebra action}\label{liealgactsect}

The Lie algebra $\su(2)$ of $\SU(2)$ is the set of traceless skew
Hermitian matrices
$$\su(2) = \left\{\left[\begin{array}{cc}
it & {u}\\
-\overline{u} & -it
\end{array}\right] \colon t\in \R, u\in \C\right\}$$
and the Lie algebra $LG=(\su(2))^n$ of the local unitary group $G=(\SU(2))^n$
is the set of $n$-tuples of matrices of this form.

A local unitary operator $g=(g_1,g_2,\ldots,g_n)$ in $G$ acts
  on a product state vector $\ket{v}=\ket{v_1}\otimes\ket{v_2}\otimes
  \cdot \otimes \ket{v_n}$ in Hilbert space $H=(\C^2)^{\otimes n}$ by
\begin{equation}
\label{liegpactionhilbert}  
g\cdot \ket{v} =
g_1\ket{v_1}\otimes g_2\ket{v_2}\otimes \cdots \otimes g_n\ket{v_n}.
\end{equation}  
The induced action on $\ket{v}$ by $X=(X_1,X_2,\ldots,X_n)$ in 
$LG$ is given by 
\begin{equation}\label{ligalgactionhilbert}
  X\cdot  \ket{v} = 
\sum_{i=1}^n \ket{v_1}\otimes\cdots\otimes \ket{v_{i-1}}\otimes
  X_i\ket{v_i}\otimes\ket{v_{i+1}}\otimes\cdots \otimes \ket{v_n}.
\end{equation}
This action extends linearly to all of Hilbert space as follows.  
Let $\ket{\psi} = \sum_I c_I|I\rangle$ be an element in
Hilbert space $H$, and let $X=(X_1,X_2,\ldots,X_n)$ be an element of
$LG$ with
$$X_k = \left[\begin{array}{cc}
it_k & {u_k}\\
-\overline{u_k} & -it_k
\end{array}\right] \hspace*{.2in}\mbox{for $1\leq k\leq n$}.$$
A straightforward calculation shows that  the action of $X$ on $\ket{\psi}$ is given by
\begin{equation}
  \label{liealgactionformula}
X\cdot \ket{\psi} = \sum_I \left(\sum_{k=1}^n (-1)^{i_k}\left[c_Iit_k + c_{I_k}\mbox{\rm
conj}^{i_k}(u_k)\right]\right)|I\rangle
\end{equation}
where $\mbox{\rm conj}^1(z)=\overline{z}$ and $\mbox{\rm conj}^0(z)={z}$.
Let $a_I$, $b_I$ denote the real and imaginary parts of the coefficient
$c_I$ in the expression for $\ket{\psi}$, and let $r_k,s_k$ denote the
real and imaginary parts of the entry $u_k$ in $X_k$.  The real and
imaginary parts of the $I$th coefficient on the right hand side of
equation~(\ref{liealgactionformula}) are the following.
\begin{eqnarray}\label{liealgactrealpart}
{\rm Re}\langle I|X|\psi\rangle &=& \sum_{k=1}^n
\left[(-1)^{i_k}(-b_It_k) + (-1)^{i_k}a_{I_k}r_k - b_{I_k}s_k\right] \\
\label{liealgactimpart}
{\rm Im}\langle I|X|\psi\rangle &=& \sum_{k=1}^n
\left[(-1)^{i_k}(a_It_k) + (-1)^{i_k}b_{I_k}r_k + a_{I_k}s_k\right]
\end{eqnarray}

Given a state $x$ in $\Proj(H)$, the isotropy Lie subalgebra $LI_x$ of the isotropy
subgroup $I_x$ is determined by the following condition.
\begin{thm}{Proposition}{liealgisoprop}
{Isotropy Lie algebra condition:}
Let $x\in \Proj(H)$ be a state and let $\ket{\psi}$ be a Hilbert space
representative for $x$.
The element $X\in LG$ is in the Lie algebra $LI_{x}$ of the
isotropy subgroup $I_{x}$ of $x$ if and only if 
$$X\cdot \ket{\psi} = i\theta \ket{\psi}$$
for some real $\theta$.
\end{thm}

With~(\ref{liealgactrealpart}) and~(\ref{liealgactimpart}),
Proposition~\ref{liealgisoprop} implies the following.
\begin{thm}{Corollary}{liealgisorealeqns}
Let $X$, $x$ and $\ket{\psi}$ be as above.
Suppose that $X$ is in $LI_x$.  Then for each multi-index $I$, we have
the following pair of equations.
\begin{eqnarray}\label{realpart}
\sum_{k=1}^n \left[(-1)^{i_k}(-b_It_k) + (-1)^{i_k}a_{I_k}r_k -
b_{I_k}s_k\right] &=& -b_I\theta\\
\label{impart}
\sum_{k=1}^n \left[(-1)^{i_k}(a_It_k) + (-1)^{i_k}b_{I_k}r_k +
a_{I_k}s_k\right] &=& a_I\theta
\end{eqnarray}
for some real number $\theta$.
\end{thm}

By adding $b_I\theta$, respectively $-a_I\theta$, to both sides of
equation~(\ref{realpart}), respectively~({\ref{impart}),  the corollary
shows that calculating the Lie algebra $LI_x$ is a matter of solving a
homogeneous real linear system of $2^{n+1}$ equations (two for each of the
$2^n$ multi-indices) in the $3n + 1$ unknowns $t_k,r_k,s_k,\theta$.  Let 
\begin{equation}
\label{bigsys}
M(t_1,r_1,s_1,t_2,r_2,s_2,\ldots,t_n,r_n,s_n,\theta) = 0
\end{equation}
denote the linear system of $2^{n+1}$ equations given
by~(\ref{realpart}) and~(\ref{impart}), so that the $2^{n+1}\times
(3n+1)$ matrix for $M$ has all entries of the form $\pm a_I,\pm b_I$.

Here is the fundamental observation which reduces the problem of orbit
dimension to finding the rank of $M$.
\begin{thm}{Proposition}{orbitdimasrank}
{Orbit dimension as rank of $M$:} Let $x$ be a state,
let $\ket{\psi}$ be a Hilbert space representative for $x$, and let $M$ be the
associated matrix constructed from the coordinates of $\ket{\psi}$ as described
above.  Then we have
$$\rank M = \dim {\cal O}_x + 1.$$
\end{thm}

\begin{proof}
We can think of $M$ as the matrix of a linear map $M\colon LG\times \R\to
\R^{2^{n+1}}$ via the identification
\begin{eqnarray*}
\R^{3n} & \leftrightarrow & LG\\
(t_1,r_1,s_1,t_2,r_2,s_2,\ldots,t_n,r_n,s_n)&\leftrightarrow&
(X_1,X_2,\ldots,X_n)   
\end{eqnarray*}
where $X_k= \twotwo{it_k}{r_k+is_k}{-r_k+is_k}{-it_k}$.
Consider a solution $(X,\theta)$ of $M(X,\theta)=0$.
Proposition~\ref{liealgisoprop} says that $\ket{\psi}$ is an eigenvector
for $X$ with eigenvalue $i\theta$, so $\theta$ is determined by $X$.
Since $X\in LI_x$ if and only if $M(X,\theta)=0$ for some $\theta$, it
follows that $\dim LI_x=\dim \ker M$.  From this we have
\begin{eqnarray*}
\dim LI_x &=&\dim \ker M\\
 &=& \mbox{number of columns of $M$} - \rank M\\
&=& 3n+1-\rank M.  
\end{eqnarray*}
Thus we have $\dim {\cal O}_x = \dim G - \dim
LI_x = 3n - (3n+1-\rank M) = \rank M -1$.
\end{proof}

Next we introduce three operators on $H$ whose purpose is to simplify
calculations (specifically, inner products of column vectors) to
establish the rank of $M$.

Let $\displaystyle A = \twotwo{i}{0}{0}{-i}$, $\displaystyle B =
\twotwo{0}{1}{-1}{0}$, and $\displaystyle C = \twotwo{0}{i}{i}{0}$
denote the standard\footnote{This basis is standard in the sense that
$A,B,C$ correspond to the truly standard basis vectors ${\mathbf
i}=(0,1,0,0)$, ${\mathbf j}=(0,0,1,0)$, ${\mathbf k}=(0,0,0,1)$ of the
pure quaternions, under the natural identification $\displaystyle
\twotwo{a}{b}{-\overline{b}}{\overline{a}} \leftrightarrow a+b{\mathbf
j}$.  In terms of the Pauli spin matrices, we have $A=i\sigma_z$,
$B=i\sigma_y$ and $C=i\sigma_x$ where $\sigma_x = \twotwo{0}{1}{1}{0}$,
$\sigma_y = \twotwo{0}{-i}{i}{0}$, and $\sigma_z =
\twotwo{1}{0}{0}{-1}$.}  basis for $\su(2)$, so that the element
$\displaystyle X = \left[\begin{array}{cc} it & {r+is}\\ -r+is & -it
\end{array}\right]$ is written $X=tA+rB+sC$ with respect to this
basis.  

Define elements $A_k,B_k,C_k$ of $LG$ for $1\leq k\leq n$ to
have $A,B,C$, respectively, in the $k$th coordinate and zero elsewhere.
\begin{eqnarray*}
A_k &=& (0,\ldots,0,\twotwo{i}{0}{0}{-i},0,\ldots,0)\\
B_k &=& (0,\ldots,0,\twotwo{0}{1}{-1}{0},0,\ldots,0)\\
C_k &=& (0,\ldots,0,\twotwo{0}{i}{i}{0},0,\ldots,0)
\end{eqnarray*}
Applying~(\ref{liealgactionformula}), we have the following.
\begin{eqnarray}
  A_k\ket{\psi} &=& \sum_{I} i(-1)^{i_k}c_{I} \ket{I}\label{akact}\\
  B_k\ket{\psi} &=& \sum_{I} (-1)^{i_k}c_{I_k} \ket{I}\label{bkact}\\
  C_k\ket{\psi} &=& \sum_{I} ic_{I_k} \ket{I}\label{ckact}
\end{eqnarray}
Simple checking shows that the complex vectors on the right hand sides of the above three
equations identify with columns of $M$ via the standard
identification~(\ref{cnidentr2n}). The
rightmost column of $M$ identifies with $-i\ket{\psi}$.  Thus we may
view $M$ as the $(3n+1)$-tuple of complex vectors
\begin{equation}\label{colsofMident}
M=(A_1\ket{\psi},B_1\ket{\psi},C_1\ket{\psi},
\ldots,A_n\ket{\psi},B_n\ket{\psi},C_n\ket{\psi},-i\ket{\psi}).
\end{equation}
It is convenient to gather the columns of $M$ into 3-tuples.  We define
the {\em triple $T_k$} to be the set of vectors
\begin{equation}
T_k=\{A_k\ket{\psi},B_k\ket{\psi},C_k\ket{\psi}\}  
\end{equation}
 for $1\leq k\leq n$.  We view the vectors both as real and also as complex
via~(\ref{cnidentr2n}). 


\setcounter{statementnumber}{0}
\section{Technical lemmas}\label{techlemmas}

In this section we present combinatorial machinery
that will be used to establish orthogonality among columns of the
matrix $M$ described in the previous section.

\begin{thm}{Lemma}{techlemma}
Let $L=(i_{jk})$ be an $\ell\times m$ matrix
with entries in $\Z_2=\{0,1\}$, and let
 $E=\displaystyle \left( (-1)^{i_{jk}}\right)$.
We view $L$ as the matrix of a $\Z_2$-linear map
$\Z_2^{m}\to\Z_2^{\ell}$ and we view $E$ as the matrix of an
$\R$-linear map. Suppose that $E$ has a nontrivial kernel.  Then either
$L$ has a nontrivial kernel or there is some $v\in \Z_2^m$
such that $L v = (1,1,\ldots,1)$. 
\end{thm}

\begin{proof}
Assume the hypotheses of the lemma.  Let $N$ be the $\ell\times m$ matrix
whose entries are all ones.  As matrices over $\R$, observe that
$E=N-2L$.

Since $E$ has integer coefficients, there is a nonzero kernel vector $v$
with integer coordinates.  Dividing by a power of 2, if necessary, we
may rescale $v$ so that the integer coordinates are not all even.  We
have $0=Ev=(N-2L)v$, so $L v = (N/2)v=(s/2)c$, where $c$ is
the column vector of all ones and $s$ is the sum of the entries in $v$.
Since $L v$ is a vector with integer entries, $L v = (s/2) c$
implies $s$ is even.  Now we can read the equation $L v = (s/2)c$
mod 2.  If $s/2=0$ mod 2, then $v$ mod 2 is a nonzero kernel vector for
$L$ since not all coordinates of $v$ are even.  If $s/2=1$ mod 2,
then $c=(1,1\ldots,1)$ is in the image of $L$.
\end{proof}

\begin{thm}{Corollary}{someconstcoordcor} 
Let $\xi_1,\xi_2,\ldots,\xi_m$ be real
numbers, not necessarily distinct, and not all of which are zero.  Let
$D_m$ be the $2^m\times 2^m$ diagonal matrix whose $r,r$ entry is
$$\sum_{i=1}^m (-1)^{r_i}\xi_i$$ where $r=(r_mr_{m-1}\ldots r_2r_1)$ is
the binary expansion of the integer $r$ in the range $0\leq r\leq
2^m-1$.  Suppose that $D_m$ has at least one zero eigenvalue.  Let
$r^1,r^2,\ldots,r^\ell$ be the row numbers of the zero eigenvalues of
$D_m$.  Then there is a nonempty set ${\cal K}=\{k_1,k_2,\ldots,k_{m'}\}$ with $1\leq
k_1<k_2<\cdots<k_{m'}\leq m$ and $m'$ even so that
$$\sum_{k\in {\cal K}} r^1_{k} = \sum_{k\in {\cal K}} r^2_{k} =\cdots=
\sum_{k\in {\cal K}} r^\ell_{k}$$
where the sums are taken mod 2.
\end{thm}

\begin{proof}
Let $\displaystyle L=\left( r^i_j\right)$ and let $\displaystyle
E=\left( (-1)^{r^i_j}\right)$.  Since $E$ kills the nonzero vector
$(\xi_1,\xi_2,\ldots,\xi_m)$, Lemma~\ref{techlemma}
applies.  If $L$ is not injective, let $v=(v_1,v_2,\ldots,v_m)$ be
a nonzero kernel vector and let $k_1,k_2,\ldots,k_{m'}$ be the indices
$i$ in the range from 1 to $m$ inclusive for which $v_i=1$.  Then the
mod 2 equation $L v=0$ yields
$$0=\sum_{k\in {\cal K}} r^1_{k} = \sum_{k\in {\cal K}} r^2_{k} =\cdots=
\sum_{k\in {\cal K}} r^\ell_{k}.$$
If there is a $v=(v_1,v_2,\ldots,v_m)$ such that $L v=(1,1,\ldots,1)$, then setting
$k_1,k_2,\ldots,k_{m'}$ to be the indices $i$ for which $v_i=1$, then we have
$$1=\sum_{k\in {\cal K}} r^1_{k} = \sum_{k\in {\cal K}} r^2_{k} =\cdots=
\sum_{k\in {\cal K}} r^\ell_{k}.$$
To see that $m'$ must be even, note that if 
$$0=\sum_{i=1}^m (-1)^{r_i}\xi_i$$
then we also have
$$0=-\sum_{i=1}^m (-1)^{r_i}\xi_i= \sum_{i=1}^m (-1)^{r_i^c}\xi_i.$$ So
if $r^1=r$ is a row number for a zero entry in $D_m$, so is $r^2=r^c$,
where $r^c$ is the binary string obtained from $r$ by complementing each
bit.  Since these two rows have opposite parity in each bit, $m'$ cannot
be odd.  This completes the proof.
\end{proof}

\begin{rem}{Definition}{consttotalparity} 
For the set ${\cal K}=\{k_1,k_2,\ldots,k_{m'}\}$
arising from zero entries in $D_m$ in row numbers $r^1,r^2,\ldots,r^l$
as in~\ref{someconstcoordcor} above, we define the {\em parity of
  ${\cal K}$} to be the common value in $\Z_2$ of the sums
$$\sum_{k\in {\cal K}} r^1_{k} = \sum_{k\in {\cal K}} r^2_{k} =\cdots=
\sum_{k\in {\cal K}} r^\ell_{k}.$$
\end{rem}

Now we are ready to establish lower bounds on
the rank of $M$ by showing that inner products of certain pairs
of columns can be arranged (via local unitary equivalence operations) to
vanish.

\setcounter{statementnumber}{0}
\section{Orthogonality Results}\label{orthressect}

Throughout this section, let
$\ket{\psi}=\sum_I c_I |I\rangle \in H$
be a Hilbert space vector, and let $M$ be the associated matrix
as defined in~\S\ref{liealgactsect}.

We make repeated use of the following elementary observation about the
relationship between complex and real inner products.  Let $u,v$ be
vectors in $\C^N$ and let $u',v'$ be the corresponding vectors in
$\R^{2N}$ given by the standard identification~(\ref{cnidentr2n}).  
The complex inner product $\inprod{u}{v}$ and the real dot product
$u'\cdot v'$ are related by
\begin{equation}\label{complexvsrealinnprod}
\real(\inprod{u}{v}) = u'\cdot v'.
\end{equation}
We shall consider complex inner products given in
Table~1 among the column vectors\footnote{For the
sake of compactness we have omitted a factor of $-i$ in the inner
products~(\ref{apsi}),~(\ref{bpsi}), and~(\ref{cpsi}).  With or without
the factor $-i$, their vanishing guarantees the orthogonality of the
rightmost column vector $-i\ket{\psi}$ of $M$ to $A_k\ket{\psi}$,
$B_k\ket{\psi}$, and $C_k\ket{\psi}$.} of $M$ given
in~(\ref{colsofMident}).

\newcounter{tmpeqnnum}
\setcounter{tmpeqnnum}{\theequation}
\setcounter{equation}{0}
\renewcommand{\theequation}{\Alph{equation}}
\begin{table}\label{colinnprodtable}
\begin{eqnarray}
\bra{\psi} A_k \ket{\psi} &=& \sum_I i(-1)^{i_k}|c_I|^2\label{apsi}\\
\bra{\psi} A_j^\dag A_k \ket{\psi} &=& \sum_I (-1)^{i_j+i_k}|c_I|^2\label{aa}\\
\bra{\psi} B_j^\dag A_k \ket{\psi} &=& \sum_I i(-1)^{i_j+i_k}\overline{c_{I_j}}c_{I}\label{ba}\\
\bra{\psi} C_j^\dag A_k \ket{\psi} &=& \sum_I  (-1)^{i_k}\overline{c_{I_j}}c_{I}\label{ca}\\
\bra{\psi} B_k \ket{\psi} &=& \sum_I (-1)^{i_k}\overline{c_I}c_{I_k}\label{bpsi}\\
\bra{\psi} A_j^\dag B_k \ket{\psi} &=& \sum_I -i(-1)^{i_j+i_k}\overline{c_I}c_{I_k}\label{ab}\\
\bra{\psi} B_j^\dag B_k \ket{\psi} &=& \sum_I (-1)^{i_j+i_k}\overline{c_{I_j}}c_{I_k}\label{bb}\\
\bra{\psi} C_j^\dag B_k \ket{\psi} &=& \sum_I  -i(-1)^{i_j}\overline{c_{I_j}}c_{I_k}\label{cb}\\
\bra{\psi} C_k \ket{\psi} &=& \sum_I i\overline{c_I}c_{I_k}\label{cpsi}\\
\bra{\psi} A_j^\dag C_k \ket{\psi} &=& \sum_I  (-1)^{i_j}\overline{c_I}c_{I_k}\label{ac}\\
\bra{\psi} B_j^\dag C_k \ket{\psi} &=& \sum_I  i(-1)^{i_j}\overline{c_{I_j}}c_{I_k}\label{bc}\\
\bra{\psi} C_j^\dag C_k \ket{\psi} &=& \sum_I \overline{c_{I_j}}c_{I_k}\label{cc}
\end{eqnarray}
\caption{Inner products of pairs of columns of $M$}
\end{table}
\setcounter{equation}{\thetmpeqnnum}
\renewcommand{\theequation}{\arabic{equation}}

Our first proposition is that each triple spans three real dimensions.
\begin{thm}{Proposition}{triplesprop}
Let  $T_k=\{A_k\ket{\psi},B_k\ket{\psi},C_k\ket{\psi}\}$ be a triple of
columns of $M$.
The three vectors in the triple are orthogonal
when viewed as real vectors.  
\end{thm}

\begin{proof}
To prove the proposition, we show that inner
products~(\ref{ab}),~(\ref{ac}), and~(\ref{bc}) in Table~1 are pure
imaginary for the case $j=k$.  First, for~(\ref{ab}), the $I$th summand
is
$$-i(-1)^{i_k+i_k}\overline{c_I}c_{I_k}=-i\overline{c_I}c_{I_k}$$  and the
$I_k$th summand is 
$$-i(-1)^{i_k+1+i_k+1}\overline{c_{I_k}}c_{I}=-i\overline{c_{I_k}}c_{I}.$$  
The sum of the $I$th and the $I_k$th summands is therefore
$-2i\real(\overline{c_I}c_{I_k})$.  By pairing the summands in this way,
we see that $\bra{\psi}A_k^\dag B_k\ket{\psi}$ is pure imaginary.  Thus
it follows from~(\ref{complexvsrealinnprod}) that
$A_k\ket{\psi},B_k\ket{\psi}$ are orthogonal as real vectors.

Next we consider~(\ref{ac}) with $j=k$.  The $I$th summand is
$(-1)^{i_k}\overline{c_I}c_{I_k}$, while the
$I_k$th summand is 
$(-1)^{i_k+1}\overline{c_{I_k}}c_{I}$.  Thus the sum of the $I$th and
$I_k$th summands is $(-1)^{i_k}2i\imag(\overline{c_I}c_{I_k})$, which is
pure imaginary, so $A_k\ket{\psi},C_k\ket{\psi}$ are orthogonal as real
vectors.

Finally we check~(\ref{bc}) for $j=k$.  In this case the $I$th summand
is  $i(-1)^{i_k}|c_{I_k}|^2$ so the inner product is pure imaginary.
Therefore $B_k\ket{\psi},C_k\ket{\psi}$ are orthogonal as real vectors.
This establishes the proposition.
\end{proof}

Next we show that a nontrivial linear dependence among the columns
$A_k\ket{\psi}$ as real vectors guarantees that certain columns among
the $B_k\ket{\psi},C_k\ket{\psi}$ are orthogonal to spans of certain
sets of triples. 

\begin{thm}{Proposition}{mainprop} 
{Main orthogonality proposition:}
Suppose that 
$$\sum_{k=1}^m\xi_{k} A_{j_k}\ket{\psi}=0$$
for some $1\leq j_1<j_2<\cdots<j_m\leq n$, $\xi_j$ real and not all zero. 
Then there is a nonempty subset $K\subseteq \{j_1,j_2,\ldots,j_{m}\}$ containing
an even number of elements
such that  $B_{k}\ket{\psi}$
and $C_{k}\ket{\psi}$ are orthogonal to $-i\ket{\psi}$ and to
$A_j\ket{\psi},B_j\ket{\psi},C_j\ket{\psi}$ for all
$k\in K$,
$j\not\in K$.
\end{thm}

\begin{proof}
Let $D_m$ be the matrix constructed from $\xi_1,\ldots,\xi_m$ as
described in the technical lemmas section.  Let $c_I$ be a
nonzero state vector coefficient.  By~(\ref{akact}), the $I$th coordinate of  $\displaystyle
\sum_{k=1}^m\xi_{k} A_{j_k}\ket{\psi}$ is $\displaystyle ic_I\sum_k
(-1)^{i_{j_k}} \xi_k$, so the hypothesis $\displaystyle
\sum_{k=1}^m\xi_{k} A_{j_k}\ket{\psi}=0$ guarantees that $D_m$ has at
least one zero eigenvalue, namely $\displaystyle \sum_k
(-1)^{i_{j_k}} \xi_k$ where $I$ is any multi-index for which $c_I\neq
0$. Therefore $\xi_1,\xi_2,\ldots,\xi_m$ and $D_m$ meet the hypothesis
of Corollary~\ref{someconstcoordcor}.

Let ${\cal K}=\{k_1,k_2,\ldots,k_{m'}\}$ be the subset of
$\{1,2,\ldots,m\}$ whose existence is guaranteed
by~\ref{someconstcoordcor} with corresponding parity $b$ as defined
in~\ref{consttotalparity}, and let
$K=\{j_{k_1},j_{k_2},\ldots,j_{k_{m'}}\}$.  The set of multi-indices of
state basis vectors $|I\rangle$ is partitioned by $K$ into two
equal-sized equivalence classes by the following equivalence relation.
\begin{equation}
  (i_1,i_2,\ldots,i_n)\sim (i'_1,i'_2,\ldots,i'_n) \Leftrightarrow
  \sum_{k\in K}i_{k} = \sum_{k\in K}i'_{k} \mod 2
\end{equation}
In words, $I\sim I'$ if the parity of the sum of bits in columns in $K$
is the same for $I$ and $I'$.   Let ${\cal P}$ be the set of
multi-indices of parity class $b$ and let ${\cal P}'$ be the opposite
parity class.

We claim that all complex inner products of the
form~(\ref{bpsi})--(\ref{cc}) in Table~1 vanish for $k\in K$ and
$j\not\in K$.  From this it follows from~(\ref{complexvsrealinnprod})
that the corresponding real dot products also vanish.  Observe that for any
$I$ for which $c_I\neq 0$ we have $\displaystyle \sum_k (-1)^{i_{j_k}}
\xi_k=0$, so $I$ is in parity class ${\cal P}$.  So if $I,J$ are
multi-indices in opposite parity classes, at least one of $c_I,c_J$ must
be zero.  If $k\in K$ and $j\not\in K$ then multi-indices $I,I_k$ are in
opposite parity classes, and also $I_j,I_k$ are in opposite parity
classes.  Since every summand in each of the inner
products~(\ref{bpsi})--(\ref{cc}) has a factor either of the form
$\overline{c_I}c_{I_k}$ or of the form $\overline{c_{I_j}}c_{I_k}$ with
$k\in K$ and $j\not\in K$, all
of the inner products vanish.

This completes the proof.
\end{proof}

\begin{thm}{Proposition}{twocommon}
Suppose that for some $1\leq l<l'\leq n$ we have 
$A_l \ket{\psi} = A_{l'} \ket{\psi}$ and 
$C_l \ket{\psi} = C_{l'} \ket{\psi}$.
Then $A_k \ket{\psi}$, $B_k \ket{\psi}$, and $C_k \ket{\psi}$
are each orthogonal to
$-i\ket{\psi}$ and to
$A_j \ket{\psi}$, $B_j \ket{\psi},C_j \ket{\psi}$
for all $k\in \{l,l'\},j\not\in\{l,l'\}$.
\end{thm}

\begin{proof}
We claim that all of the complex (and hence also real,
by~(\ref{complexvsrealinnprod})) inner products~(\ref{apsi})--(\ref{cc})
vanish for $k\in \{l,l'\}$ and $j\not\in\{l,l'\}$.  We begin by
applying Proposition~\ref{mainprop} to the hypothesis
$A_l\ket{\psi}=A_{l'}\ket{\psi}$.  In the notation of~\ref{mainprop}
we have $m=2$ and therefore also $m'=2$ since $m'$ is an even number in
the range $0<m'\leq m$, so $K=\{l,l'\}$.  Thus we have
from~\ref{mainprop} that $B_k\ket{\psi}$ and $C_k\ket{\psi}$ are
orthogonal to $-i\ket{\psi}$ and to $A_j \ket{\psi}$, $B_j
\ket{\psi},C_j \ket{\psi}$ for all $k \in \{l,l'\},j\not\in\{l,l'\}$.

It remains to be shown that $A_l\ket{\psi},A_{l'}\ket{\psi}$
are also orthogonal to $-i\ket{\psi}$ and to $A_j \ket{\psi}$, $B_j
\ket{\psi},C_j \ket{\psi}$ for all $j \not\in \{l,l'\}$.

The hypothesis $C_l\ket{\psi}=C_{l'}\ket{\psi}$ implies that
$c_{I_l}=c_{I_{l'}}$, or equivalently, that $c_{I}=c_{I_{ll'}}$ for all
$I$.  This implies that summands of the inner
products~(\ref{apsi})--(\ref{ca}) cancel in pairs for
$k\in\{l,l'\},j\not\in\{l,l'\}$, as follows.  
The $I$th summand of~(\ref{apsi}) is $i(-1)^{i_k}|c_I|^2$ and the
$I_{ll'}$th summand is $i(-1)^{i_k+1}|c_I|^2$.
The $I$th summand of~(\ref{aa}) is $(-1)^{i_j+i_k}|c_I|^2$ and the
$I_{ll'}$th summand is $(-1)^{i_j+i_k+1}|c_I|^2$.
The $I$th summand of~(\ref{ba}) is $i(-1)^{i_j+i_k}\overline{c_{I_j}}c_I$ and the
$I_{ll'}$th summand is $i(-1)^{i_j+i_k+1}\overline{c_{I_j}}c_I$.
The $I$th summand of~(\ref{ca}) is $(-1)^{i_k}\overline{c_{I_j}}c_I$ and the
$I_{ll'}$th summand is $(-1)^{i_k+1}\overline{c_{I_j}}c_I$.

This completes the proof.
\end{proof}

\setcounter{statementnumber}{0}
\section{Local unitary adjustment}\label{orthgensect}

In this section we adapt the orthogonality results of the previous
section to hypotheses involving more general linear dependencies.

Let us write $\langle T_{i_1},T_{i_2},\ldots,T_{i_r}\rangle$ to denote
the subspace of the (real) column space of $M$ spanned by the vectors in the
triples $T_{i_1},\ldots,T_{i_r}$ viewed as real vectors.

\begin{thm}{Proposition}{mainpropgen}
{Main orthogonality proposition generalized:} Suppose that
$$\dim\langle T_{j_1},T_{j_2},\ldots,T_{j_m}\rangle<3m$$ for some
$1\leq j_1<j_2<\cdots<j_m\leq n$.  Then there is a nonempty subset
$K\subseteq \{j_1,j_2,\ldots,j_m\}$ containing an even number of elements such
that there are two orthogonal vectors $\ket{\zeta_k},\ket{\eta_k}$ in
$\langle T_k \rangle$, both of which are orthogonal to $-i\ket{\psi}$,
$A_j\ket{\psi}$, $B_j\ket{\psi}$ and to $C_j\ket{\psi}$ for all $k\in K,
j\not \in K$.
\end{thm}

\begin{proof}
Let us write the linear dependency as a relation
$$\sum_{i=1}^m\xi_i \ket{\phi_i}=0$$ where $\xi_i$ is real,
$\ket{\phi_i}$ lies
in $\langle T_{j_i}\rangle $, 
$\inprod{\phi_i}{\phi_i}=\inprod{\psi}{\psi}$ for $1\leq i\leq m$, and
not all the $\xi_i$ are zero.
Write each $\ket{\phi_i}$ as a linear combination
$$\ket{\phi_i} = \alpha_i A_{j_i}\ket{\psi} + \beta_i B_{j_i}\ket{\psi}
+ \gamma_i C_{j_i}\ket{\psi}$$ with $\alpha_i$, $\beta_i$ and $\gamma_i$
real.  Let $R_i\in \SO(\su(2))$
be such that
$$R_i(A) = \alpha_i A + \beta_i B + \gamma_i C.$$ Since the adjoint
representation $\mbox{\rm Ad}\colon \SU(2)\to \SO(\su(2))$ is
surjective, we can choose $U_{j_i}\in \SU(2)$ such that $\mbox{\rm
Ad}(U_{j_i}^\dag) = R_i$, that is, $U_{j_i}^\dag X U_{j_i} = R_i X$ for
all $X\in \su(2)$.  For
$j\not\in \{j_1,j_2,\ldots,j_m\}$, set $U_j$ equal to the identity.
Finally, let $U\in G=\SU(2)^n$ be $U=\prod_{i=1}^n U_i$.

Now observe that 
$$\sum_{i=1}^m \xi_i (U^\dag A_{j_i} U)\ket{\psi} = \sum_{i=1}^m \xi_i
\ket{\phi_i} = 0.$$ Applying $U$ to both sides, we get
$$\sum_{i=1}^m \xi_i A_{j_i}(U\ket{\psi})=0.$$ Let $M'$ be the matrix
for the state vector $U\ket{\psi}$.  Applying the main orthogonality
proposition~\ref{mainprop} to $M'$, we get that $B_{k}(U\ket{\psi})$,
$C_{k}(U\ket{\psi})$ are orthogonal to $U\ket{\psi}$ and to
$A_jU\ket{\psi},B_jU\ket{\psi},C_jU\ket{\psi}$ for $k\in K,j\not\in K$.
Now set
\begin{eqnarray*}
  \ket{\zeta_k} &=& U^\dag B_k U\ket{\psi}\\
  \ket{\eta_k} &=& U^\dag C_k U\ket{\psi}
\end{eqnarray*}
for $k\in K$.  Since $U$ is unitary, we have that
$\ket{\zeta_k},\ket{\eta_k}$ are orthogonal to $U^\dag
U\ket{\psi}=\ket{\psi}$ and to $U^\dag A_jU\ket{\psi},U^\dag
B_jU\ket{\psi},U^\dag C_jU\ket{\psi}$ for $k\in K,j\not\in K$.  Since
the three vectors
$U^\dag A_jU\ket{\psi},U^\dag B_jU\ket{\psi},U^\dag C_jU\ket{\psi}$ have
the same span as $A_j\ket{\psi},B_j\ket{\psi},C_j\ket{\psi}$ for all $j$, the
proposition is established.
\end{proof}

\begin{thm}{Proposition}{twocommongen}
{Generalization of~\ref{twocommon}:} Suppose that
$\dim \langle T_l,T_{l'}\rangle \leq 4$ for some $1\leq l<l'\leq n$.
Then $A_k \ket{\psi}$, $B_k \ket{\psi}$, and $C_k \ket{\psi}$ are each
orthogonal to $-i\ket{\psi}$ and to $A_j \ket{\psi}$, $B_j \ket{\psi},C_j
\ket{\psi}$ for all $k\in \{l,l'\},j\not\in\{l,l'\}$.
\end{thm}

\begin{proof}
The proof is very similar to the proof of~\ref{mainpropgen}.

Since $\dim\langle T_l,T_{l'}\rangle \leq 4$, the dimension of the
intersection $\langle T_l\rangle \cap \langle T_{l'}\rangle$ is at least
two.  Choose orthogonal vectors $\ket{\phi},\ket{\phi'}$ in $\langle
T_l\rangle \cap \langle T_{l'}\rangle$ with
$\inprod{\phi}{\phi}=\inprod{\phi'}{\phi'}=\inprod{\psi}{\psi}$.  Write
linear combinations
\begin{eqnarray*}
\ket{\phi} &=& \alpha_k A_{k}\ket{\psi} + \beta_k B_{k}\ket{\psi}
+ \gamma_k C_{k}\ket{\psi}\\
\ket{\phi'} &=& \alpha_k' A_{k}\ket{\psi} + \beta_k' B_{k}\ket{\psi}
+ \gamma_k' C_{k}\ket{\psi}
\end{eqnarray*}
and let $R_k\in \SO(\su(2))$
be such that
\begin{eqnarray*}
R_k(A) = \alpha_k A + \beta_k B + \gamma_k C\\  
R_k(C) = \alpha_k' A + \beta_k' B + \gamma_k' C
\end{eqnarray*}
for $k\in\{l,l'\}$.

Since the adjoint
representation $\mbox{\rm Ad}\colon \SU(2)\to \SO(\su(2))$ is
surjective, we can choose $U_{k}\in \SU(2)$ such that $\mbox{\rm
Ad}(U_{k}^\dag) = R_k$, that is, $U_{k}^\dag X U_{k} = R_k X$ for
all $X\in \su(2)$.  For
$j\not\in \{l,l'\}$, set $U_j$ equal to the identity.
Finally, let $U\in G=\SU(2)^n$ be $U=\prod_{i=1}^n U_i$.

Now observe that
\begin{eqnarray*}
  U^\dag A_l U\ket{\psi} &=& \ket{\phi}= U^\dag A_{l'}U\ket{\psi}\\
  U^\dag C_l U\ket{\psi} &=& \ket{\phi'}= U^\dag C_{l'}U\ket{\psi}
\end{eqnarray*}
Applying $U$ to both sides, we get
\begin{eqnarray*}
  A_l U\ket{\psi} &=&  A_{l'}U\ket{\psi}\\
  C_l U\ket{\psi} &=&  C_{l'}U\ket{\psi}
\end{eqnarray*}

Let $M'$ be the matrix
for the state vector $U\ket{\psi}$.  Applying~\ref{twocommon} to $M'$, we get that
$A_{k}(U\ket{\psi})$, $B_{k}(U\ket{\psi})$, $C_{k}(U\ket{\psi})$ are orthogonal to
$U\ket{\psi}$ and to $A_jU\ket{\psi},B_jU\ket{\psi},C_jU\ket{\psi}$ for
$k\in \{l,l'\},j\not\in \{l,l'\}$.  
Since $U$ is unitary, we have that $U^\dag A_kU\ket{\psi},U^\dag
B_kU\ket{\psi},U^\dag C_kU\ket{\psi}$
 are orthogonal to $U^\dag
U\ket{\psi}=\ket{\psi}$ and to $U^\dag A_jU\ket{\psi},U^\dag
B_jU\ket{\psi},U^\dag C_jU\ket{\psi}$ for $k\in \{l,l'\},j\not\in \{l,l'\}$.  Since
the three vectors
$U^\dag A_jU\ket{\psi},U^\dag B_jU\ket{\psi},U^\dag C_jU\ket{\psi}$ have
the same span as $A_j\ket{\psi},B_j\ket{\psi},C_j\ket{\psi}$ for all $j$, the
proposition is established.
\end{proof}

\setcounter{statementnumber}{0}
\section{Minimum Orbit Theorem}\label{minorbthm}

\begin{thm}{Theorem}{minorbthmstmnt}
{Minimum orbit dimension:} For the local unitary group
action on state space for $n$ qubits, the smallest orbit dimension is
$$\min\{\dim {\cal O}_x\colon x\in \Proj(H)\} = \left\{
\begin{array}{cc}
\frac{3n}{2} & \mbox{$n$ even}\\
\frac{3n+1}{2} & \mbox{$n$ odd}
\end{array}\right..$$
\end{thm}

We begin the proof by exhibiting a state for which the claimed minimum
dimension is realized.

Let $\ket{s}=|01\rangle - |10 \rangle$ be a Hilbert space representative
  of the singlet state, let
  $X=\twotwo{it}{u}{-\overline{u}}{-it}$ be an element of $\su(2)$.  We have
$$(X,X)\cdot \ket{s} = 0$$
so $(X,X)$ is in the isotropy Lie algebra
of the state represented by $\ket{s}$.

From this it follows that $(X_1,X_1,X_2,X_2,\ldots,X_k,X_k)$ stabilizes
the $2k$-qubit state $x$ represented by $\underbrace{\ket{s}\otimes \ket{s}\otimes \cdots \otimes
  \ket{s}}_{\mbox{\small $k$ copies}}$ for all
$X_1,X_2,\ldots, X_k$ in $\su(2)$.  Therefore $LI_x$ has dimension at least
$3k=3n/2$ for $n=2k$, and therefore $\dim{\cal O}_x\leq 3n-3n/2=3n/2$
for $n$ even.

Observe that $(X_1,X_1,X_2,X_2,\ldots,X_k,X_k,\left[\begin{array}{cc}it
    & 0\\0 & -it\end{array}\right])$ stabilizes the $(2k+1)$-qubit state
    $x$ represented by $\ket{s}\otimes \ket{s}\otimes \cdots \otimes
    \ket{s}\otimes |0\rangle$ (by a phase factor), so the dimension of
    $LI_x$ is at least $3k+1=(3n-1)/2$ for $n=2k+1$, and therefore
    $\dim{\cal O}_x\leq 3n-(3n-1)/2=(3n +1)/2$ for $n$ odd.

These calculations establish that
\begin{equation}\label{upperboundminorbdim}
\min\{\dim {\cal O}_x\colon x\in \Proj(H)\} \leq \left\{
\begin{array}{cc}
\frac{3n}{2} & \mbox{$n$ even}\\
\frac{3n+1}{2} & \mbox{$n$ odd}
\end{array}\right..
\end{equation}

Next we show that this bound on orbit dimension is sharp by establishing
a lower bound for the rank of $M$.  From~\ref{minrankM} below, the desired
lower bound for the minimum orbit dimension follows immediately
from~\ref{orbitdimasrank}. 

\begin{thm}{Proposition}{minrankM}
{Minimum rank of $M$:} Let $x$ be a state for a system of $n$
qubits, let $\ket{\psi}$ be a Hilbert space representative for $x$, and
let $M$ be the real matrix associated to $\ket{\psi}$ as defined
in~\S\ref{liealgactsect}.  We have
$$\rank M \geq \left\{
\begin{array}{cc}
\frac{3n}{2} + 1 & \mbox{$n$ even}\\
\frac{3n+1}{2} + 1 & \mbox{$n$ odd}
\end{array}\right..$$
\end{thm}

\begin{proof}
Let ${\cal C}=\{A_1\ket{\psi},B_1\ket{\psi},C_1\ket{\psi},
\ldots,A_n\ket{\psi},B_n\ket{\psi},C_n\ket{\psi},-i\ket{\psi}\}$ denote
the set of columns of $M$.  For a subset ${\cal S}\subseteq {\cal C}$,
let $\langle {\cal S}\rangle$ denote the real span of the column vectors
contained in ${\cal S}$.
Let ${\cal S}_0$ be a subset of ${\cal C}$ which is the union of some
number $p$ of triples, and is maximal with respect 
to the property that $\langle {\cal S}_0 \rangle$ contains a subspace
$W$ for which
\begin{enumerate}
\item[(i)] \label{propa} $\displaystyle \dim W
\geq  \left\{ \begin{array}{cc} \frac{3p}{2} & \mbox{$p$ even}\\
\frac{3p+1}{2} & \mbox{$p$ odd}\end{array}\right.$,\hspace*{.25in}
and
\item[(ii)] \label{propb} $W \perp \langle{\cal
C}\setminus{\cal S}_0 \rangle$.
\end{enumerate}

We separate the argument into cases.  We show that in every case,
either~\ref{minrankM} holds or we can derive a contradiction by
constructing a superset ${\cal S}_1$ of ${\cal S}_0$ which is the union
of some number $p'>p$ of
triples and which contains a subspace $W'$ satisfying properties~(i) and
~(ii) with $p'$ in place of $p$.  The construction of ${\cal S}_1$
violates the maximality of ${\cal S}_0$ and therefore rules out the case
in question.

Case 1: Suppose that $p=n$, so that ${\cal C}\setminus{\cal
S}_0=\{-i\ket{\psi}\}$.  Then property~(ii) guarantees that $\rank M
\geq \dim W +1$, so property~(i) guarantees that~\ref{minrankM} holds.

Case 2: Suppose that $p<n$ and that the remaining triples
$T_{j_1},T_{j_2},\ldots,T_{j_{n-p}}$ in ${\cal C}\setminus {\cal S}_0$
have the maximum possible span, that is, 
$$\dim \langle T_{j_1},T_{j_2},\ldots,T_{j_{n-p}}\rangle =3(n-p).$$
Properties~(i) and~(ii) imply that 
\begin{eqnarray*}
\rank M &\geq& \dim W + \dim \langle{\cal C}\setminus{\cal S}_0 \rangle
\\
&\geq& \frac{3p}{2} + 3(n-p)\\
&=& \frac{6n-3p}{2}\\
&\geq& \frac{6n - (3n-3)}{2}\hspace*{.2in}\mbox{(since $p\leq n-1$)}\\
&=& \frac{3n+3}{2}\\
&=& \frac {3n+1}{2} + 1
\end{eqnarray*}
and so~\ref{minrankM} holds.
Note that if $p=n-1$, the hypothesis of full span is met
by~\ref{triplesprop}.  Therefore in the remaining cases we need only
consider $p\leq n-2$.

Case 3: Suppose $p\leq n-2$ and that there is a
pair of triples $T_l,T_{l'}$ in ${\cal C}\setminus{\cal S}_0$ with
$1\leq l<l'\leq n$ such that
$\dim\langle T_l,T_{l'}\rangle \leq 4$.   
Let ${\cal S}_1 = {\cal S}_0 \cup T_l\cup T_{l'}$, let
$p'=p+2$, and let $W'=W \oplus \langle T_l\cup T_{l'} \rangle$, where
``$\oplus$'' denotes the orthogonal direct sum.  That the sum is
orthogonal is guaranteed by property~(ii) for $W$.
Proposition~\ref{twocommongen} implies that property~(ii) also holds
for the pair $({\cal S}_1,W')$ and that
$\dim W' \geq \dim W + 3$.  It follows that if $p$ is even, so is $p'$ and we have
$$\dim W' \geq \frac{3p}{2} +3 = \frac{3p+6}{2} = \frac{3(p+2)}{2} =
\frac{3p'}{2}$$
and similarly if $p$ and $p'$ are odd we have
$$\dim W' \geq \frac{3p+1}{2} +3 = \frac{3p'+1}{2}$$
so $({\cal S}_1, W')$ satisfies property~(i).  Thus ${\cal S}_1$ violates the
maximality of ${\cal S}_0$, so we conclude that the hypothesis of case~3
is impossible.

Case 4: Suppose $p\leq n-2$ and that there is a pair of triples
$T_l,T_{l'}$ in ${\cal C}\setminus{\cal S}_0$ with $1\leq l<l'\leq n$
such that $\dim\langle T_l,T_{l'}\rangle =5$.
Applying~\ref{mainpropgen} we have four vectors
$$\ket{\zeta_l},\ket{\eta_l}\in \langle T_l\rangle, \hspace*{.25in}
\ket{\zeta_{l'}},\ket{\eta_{l'}}\in \langle T_{l'} \rangle$$ 
which must span at least three dimensions, so once again ${\cal S}_1 =
{\cal S}_0 \cup T_l\cup T_{l'}$ with the subspace 
$$W'=W\oplus \langle \ket{\zeta_l},\ket{\eta_l}, \ket{\zeta_{l'}},\ket{\eta_{l'}}\rangle $$
violates the maximality of ${\cal S}_0$.  We conclude that the
hypothesis of case~4 is impossible.  

Case 5: The only remaining possibility is that $p\leq n-3$.
Let ${\cal T}=\{T_{j_1},T_{j_2},\ldots,T_{j_m}\}$ be a set of
triples in ${\cal C}\setminus {\cal S}_0$ with $m\geq 3$ minimal with
respect to the property 
$$\dim \langle T_{j_1},T_{j_2},\ldots,T_{j_m}\rangle < 3m.$$
Applying~\ref{mainpropgen} we have two vectors
$$\ket{\zeta_{k}},\ket{\eta_{k}}\in \langle T_{k}\rangle$$
for each of the $m'\geq 2$ elements $k\in K$.  Let 
$${\cal S}_1 = {\cal S}_0 \cup \left(\bigcup_{k\in K} T_k\right),$$
let $p'=p+m'$, and let 
$$W'=W \oplus \langle \{\ket{\zeta_{k}},\ket{\eta_{k}}\}_{k\in
  K}\rangle.$$ Note that property~(ii) holds for $({\cal S}_1,W')$.  If
  $m'<m$, then the $2m'$ vectors in
  $\{\ket{\zeta_{k}},\ket{\eta_{k}}\}_{k\in K} $ are independent by the
  minimality of ${\cal T}$, so we have
$$\dim W' \geq \dim W + 2m' \geq \frac{3p}{2} + 2m' =
  \frac{3p'+m'}{2}\geq \frac{3p'+1}{2} $$
so property~(i) holds for $({\cal S}_1,W')$, but this contradicts the
  maximality of ${\cal S}_0$.  Finally, if $m'=m$, then $m\geq 4$ (since
  $m'$ is even) and at least $2(m-1)$ of
  the vectors in $\{\ket{\zeta_{k}},\ket{\eta_{k}}\}_{k\in K} $ must be
  independent, again by the minimality of ${\cal T}$.  If $p$ is even,
  then $p'=p+m$ is also even and we have
$$\dim W' \geq \dim W + 2(m-1) \geq \frac{3p}{2} + 2(m-1) =
  \frac{3p'+m-4}{2}\geq \frac{3p'}{2}.$$  
If $p$ is odd, then $p'=p+m$ is odd and we have
$$\dim W' \geq \dim W + 2(m-1) \geq \frac{3p+1}{2} + 2(m-1) =
  \frac{3p'+m-3}{2}\geq \frac{3p'+1}{2}.$$  
Thus ${\cal S}_1$ with the subspace $W'$ violates the maximality
  of ${\cal S}_0$.  We conclude that the hypothesis of case~5 is impossible.

Having exhausted all possible cases, this completes the proof
of~\ref{minrankM}, and hence of Theorem~\ref{minorbthmstmnt}.
\end{proof}

\section*{Acknowledgments}
We are grateful for funding support from Lebanon Valley College for the
initiation of this project in 2002--2003.  S.N.W. thanks the Research
Corporation for their support.

\bibliographystyle{unsrt}


\end{document}